\documentstyle[prl,epsf,eqsecnum,aps]{revtex}

\begin{document}

\title{Kaluza-Klein Induced Gravity Inflation}
\author{W. F. Kao \thanks{email:wfgore@cc.nctu.edu.tw} \\
Institute of Physics, Chiao Tung University, Hsin Chu, Taiwan
\vspace{1cm} \\ (Revised April, 2000)}


\maketitle

\begin{minipage}[t]{17.5cm}
\begin{abstract}
A D-dimensional induced gravity theory is studied carefully in
a $4 + (D-4)$ dimensional
Friedmann-Robertson-Walker space-time.
We try to extract information of the symmetry breaking potential
in search of an inflationary solution with non-expanding internal-space.
We find that the induced gravity model imposes  strong constraints on 
the form of symmetry breaking potential in order to generate an acceptable inflationary universe.
These constraints are analyzed carefully in this paper.
\end{abstract}

PACS numbers:  04.50.+h, 98.80Cq, 02.40.-k

 \end{minipage}

\section{Introduction}
Scale invariant model explains the origin of the scale
parameters such as
gravitational constant, cosmological constant as well as the masses for the
fermion fields. Accordingly, all dimensionful
parameters in Einstein Lagrangian are replaced by appropriate scalar
measuring field with proper power according to their conformal dimensions.

Scale invariance also appears to be very important in various branches
of physics such as QCD \cite{ryder} and many other inflationary models
\cite{zee,ni,acc}. 
Local scale symmetry has also been suggested to be related to the missing Higgs
problem in electro-weak theory \cite{CK} as well as many other research interests 
\cite{smolin,Lee}. It is also argued that scale invariant effective theory
has to do with physics near the fixed points of renormalization group trajectory
\cite{smolin}.

On the other hand, higher dimensional Kaluza-Klein theory 
\cite{kim,visser, green}  has been a focus of research interest for a long time.
In addition, Kaluza-Klein theory should be related to the evolution of our early universe if the compactification process is completed during the early stage of the universe. 
Hence one is naturally lead to the question
whether scale invariant effective action is manifest before dimensional
reduction process takes place. Therefore we propose to study
the effect of a $D$ dimensional induced gravity in 
the very early universe. 

One notes that there have been studies
based on a $D$-dimensional Friedmann-Robertson-Walker (DFRW) metric
\begin{equation}
 ds^{2} \equiv \hat{g}_{MN}dz^{M} dz^{N} = - dt^{2} + a^{2} (t) \left[
 \frac{1}{1 - kr^{2} } dr^{2} + r^{2} d\Omega_{D-1}  \right]
\label{metricd}
\end{equation}
in search of a physically acceptable low energy effective theory 
\cite{green,renor,equiv,ms,kal}.
Here $
d\Omega_{n}$ denotes the n-dimensional solid angle with appropriate
angular coordinates.
Note, however, that the internal and external spaces will inflate all together if $a(t)$ undergoes an exponentially expanding process under the
DFRW metric.
Hence it would be interesting to see if a more general
$4$+$d$ dimensional FRW (4dFRW) space is capable of inducing inflating
external-space and contracting internal-space at the same time. 
Here $d \equiv D-4$ denotes the dimension of the internal space.
This kind of generalization is in fact a dimensional reduction
from $M^D \to R^1 \times F^3 \times F^d$ with $F^n$ denoting the
$n$-dimensional maximally symmetric space \cite{weinberg1}.

In this paper, we will show that an induced inflationary solution with 
expanding external-space and contracting 
internal-space will require a very special symmetry breaking scalar
potential. The explicit form of the 
symmetry breaking potential required by the expanding solution with expanding 
external-space and constant internal-space 
will also be solved. In addition, we will also show that the
most favorable solution appears to be the same as the result of
DFRW space for the induced conventional-$\phi^4$ model.
Our result indicates that the compactification process must have been completed
before the inflationary process unless the symmetry breaking potential takes an unordinary form.

This paper will be organized as follows:
(i) In section II, we will introduce the $D$-dimensional induced gravity
theory.
The $D$-dimensional equations of motion will also be
compactified into $4+d$ dimensional in this section.
(ii) We will also solve the equations of motion for an inflationary solution in the
limit of slow-rollover approach in section III. It will be shown first that 
the existence of a solution with expanding external-space and 
contracting internal-space 
will impose a number of constraints on the coupled symmetry breaking potential.
In particular, solution with expanding external-space and constant internal-space
is also solved in search of the possible candidate for
the scalar potential.
(iii) the conventional $\phi^4$ model with a coupled SSB $\phi^4$ 
potential is solved and analyzed in section IV. 
(iv) Finally, conclusions are drawn in section V. 

\section{Induced Gravity in $D$ Dimension}
In this paper we will consider the following induced gravity action :
\begin{equation}
S = - \int d^{D}z \sqrt{\hat{g}}\ \Big[\:
\frac{1}{2}\epsilon\phi^{2} \hat{R}
+     \frac{1}{2}\partial^{M}\phi\partial_{M}\phi +
 V(\phi) \: \Big] .
\label{action}
\end{equation}
The scalar field $\phi$ in (\ref{action}) is the measuring field designed to
replace the dimensionful
Newtonian constant as appeared in the four dimensional induced gravity models.
One should replace all dimensionful coupling constants with appropriate scalar
fields according to their dimensions.
After this replacement, one can show that, apart from
a possible symmetry breaking potential $V$,
the action
(\ref{action}) is invariant under the following scale transformation: $g_{MN}
\to \Lambda^{2}g_{MN}$ , $\phi \to \Lambda^{(1-D/2)}\phi$.

We will denote $D=4+d$ from now on in this paper. Here $d=D-4$ is the
dimension of the internal space
in the Kaluza-Klein theory we are going to study in this paper.
We will also use a hat notation, $\hat{R}_{MN}$, to represent the physical
variables in $D$-dimension and non-hatted variable, $R_{ab}$, will represent
the same physical variables evaluated solely in the $4$-dimensional physical
external-space. Barred notation, $\bar{R}_{mn}$, will also be employed to denote the
same physical variables evaluated in $d$-dimensional internal-space.
Note that by {\it the same physical variables} we meant that they
are defined by the same notation except the metric is replaced by
the appropriate metric defined in its own space.
Furthermore, we will use capital indices $M, N, \cdots$ $(=0,1,2, \cdots, D-1)$
to denote
the $D$ dimensional space-time indices. Also lower case Latin indices from
the beginning ($a,b,c,
\cdots$) of the alphabet will denote the four dimensional space
time indices $(a,b,c = 0,1,2,3)$.
In addition, $i,j,k,l (=1,2,3)$ labels the spatial $3$-manifold.
Finally, we will
use lower case Latin indices from the middle ($m,n, \cdots$) of the alphabet to
label the
$d$-dimensional compactified internal-space.

Note that the Kaluza-Klein dimensional reduction process we will adopt
in this paper is the following $4+d$ dimensional 
Friedmann-Robertson-Walker metric (4dFRW) \cite{weinberg1,meyer}
\begin{equation}
 ds^{2} \equiv \hat{g}_{MN}dz^{M} dz^{N} = - dt^{2} + a^{2} (t) h_{ij} (x)
dx^{i} dx^{j}        + b^{2} (t) \bar{h}_{mn} (y) dy^{m} dy^{n}  \label{metric} .
\end{equation}
Here $
h_{ij} dx^{i}dx^{j} \equiv (1 - k_{1}r^{2})^{-1} dr^{2} + r^{2}
d\Omega_{3} $ and 
$\bar{h}_{mn} dy^{m}dy^{n} \equiv (1 - k_{2}s^{2})^{-1}  ds^{2} + s^{2}
d\Omega_{d}$ with $k_1,\,k_2= 0, \pm 1$ denoting the signature of the external space
and internal space respectively.

Note that if we adopt the compactification ansatz $\phi(z) =\phi(x)
\kappa^{d\over 2}$, the compactified $4$-dimensional effective Einstein
action, except the SSB $\phi^4$ potential term, will remain scale invariant
under the $4$-dimensional scale transformation:
$g_{ab} (x)
\to \Lambda(x)^{2}g_{ab}(x)$ , $\phi(x) \to \Lambda(x)^{(2-D)/2}\phi(x)$.
This shows that this is a consistent and scale-invariant way to carry out the compactification
process. Note also that $\kappa$ is a dimension one constant parameter such
that $\int d^{d}y \kappa^{d}$ is dimensionless and will be set as $1$
for latter convenience.
Note that
we will also use the same $\phi$ notation for $\phi(z)$ and $\phi(x)$
for convenience. 

The equations of motion can be obtained from varying the action (\ref{action})
with respect
to $\phi$ and $\hat{g}_{MN}$ respectively. As a result, one has:
\begin{eqnarray}
\epsilon\phi \hat{R} - D_{M}\partial^{M}\phi
       + \frac{\partial}{\partial\phi} V(\phi) &=& 0,  \label{eqn1} \\
\epsilon \phi^{2} \hat{G}_{MN} &=& \epsilon ( D_{M} \partial_{N}
        - \hat{g}_{MN} D^{P} \partial_{P} ) \phi^{2}
      +  \hat{T}_{MN}^{\phi}  \: .
\label{eqn3}
\end{eqnarray}
Here \( \hat{G}_{MN} \equiv \frac{1}{2} \hat{R} \hat{g}_{MN} - \hat{R}_{MN} \)
defines the Einstein tensor. Moreover, \(
\hat{T}^{MN}_{\phi} \equiv \partial^{M} \phi \partial^{N} \phi
  - \hat{g}^{MN} [   \frac{1}{2} \partial^{P} \phi \partial_{P} \phi + V(\phi)
] \) is the energy momentum tensor associated with $\phi$.
Furthermore, the curvature
tensor $\hat{R}_{MNOP}$ is defined by \( [ D_{M},D_{N} ]\hat{A}_{O} =
\hat{R}^{P}_{ONM}\hat{A}_{P} \).
In addition, the Ricci tensor and scalar curvature are defined
by $\hat{R}_{ON}\equiv \hat{R}^{P}_{ONP}$ and
$\hat{R}\equiv \hat{R}_{ON} \hat{g}^{ON}$ respectively.

For latter convenience,
we will define $\phi^{2} \equiv e^{\varphi}$, $a \equiv e^{\alpha}$, $b
\equiv e^{\beta}$,  $V (\phi) \equiv U (\varphi)$ throughout this paper.
Hence one can bring (\ref{eqn1}) and (\ref{eqn3}) into a more
comprehensive form
in terms of the new variables and parameters
defined earlier. Indeed, (\ref{eqn1}) and (\ref{eqn3}) can be written as
\begin{equation}
\hat{G}_{MN} = D_{M}\partial_{N}\varphi +
\partial_{M}\varphi\partial_{N}\varphi
       - \hat{g}_{MN}( D^{P}\partial_{P}\varphi +
\partial^{P}\varphi\partial_{P}\varphi)
       - \hat{T}^{\varphi}_{MN}  \: ,
\label{gmnd}
\end{equation}
\begin{equation}
\hat{R} = {1 \over 4\epsilon} ( \partial^{M}\varphi \partial_{M}\varphi
        + 2 D^{P}\partial_{P}\varphi )
        - {8 \over \epsilon} e^{-\varphi} \frac{\partial U(\varphi)}{\partial\varphi} \: .
\label{phid}
\end{equation}
Hence one has
\begin{eqnarray}
\hat{R} &=& {1 \over 4 \epsilon} (\partial_a \varphi \partial^a \varphi + D_a \partial^a
\varphi)  -
 \frac{2}{\epsilon}  e^{-\varphi} \partial_\varphi U(\varphi) ,
\label{phi2} \\
\hat{G}_{ab} &=& ( \partial_{a}\varphi \partial_{b}\varphi +
D_{a}\partial_{b}\varphi )
       -  g_{ab} ( \partial^{c}\varphi \partial_{c}\varphi +
D^{P}\partial_{P}\varphi )        - T_{ab}^{\varphi}  \:  ,
\label{gab2} \\
\hat{G}_{mn} &=& D_{m}\partial_{n}\varphi - \bar{g}_{mn} ( \partial^{c}\varphi
\partial_{c}\varphi
       + D^{P}\partial_{P}\varphi ) - \bar{T}_{mn}^{\varphi}
\: . \label{gmn2}
\end{eqnarray}
Here we have defined the generalized energy momentum tensor for $\varphi$ and
$\hat{T}^{\varphi}_{MN}$ as:
\begin{equation}
\hat{T}_{MN}^{\varphi} = {1 \over 4 \epsilon} ( \frac{1}{2} \hat{g}_{MN}
\partial_{P}\varphi \partial^{P}\varphi
                 - \partial_{M}\varphi \partial_{N}\varphi )
                 + \frac{V}{\epsilon} e^{-\varphi} \hat{g}_{MN} \: .
\end{equation}
Therefore, one has
\begin{eqnarray}
T_{ab}^{\varphi}  &=& {1 \over 4 \epsilon} ( \frac{1}{2} g_{ab} \partial^{c}\varphi
\partial_{c}\varphi
                 - \partial_{a}\varphi \partial_{b}\varphi )
                 + {1 \over  \epsilon} e^{-\varphi} U(\varphi) g_{ab} \: , \\
\bar{T}_{mn}^{\varphi} &=& \frac{1}{2} {1 \over 4 \epsilon} \bar{g}_{mn}
\partial^{c}\varphi \partial_{c}\varphi
                 +  {1 \over  \epsilon} e^{-\varphi} U(\varphi) \bar{g}_{mn} \: ,
\end{eqnarray}
respectively.
In addition, with the compactified metric
\begin{equation}
ds^{2} \equiv \hat{g}_{MN} dz^{M}dz^{N} = g_{ab}(z) dx^{a}dx^{b} +
\bar{g}_{mn}(z)dy^{m}dy^{n} \: ,         \label{metricD}
\end{equation}
and by setting $\phi(z)=\phi(x)$,
one can derive the following compactified identities for the
curvature terms: 
\begin{eqnarray}
\hat{R} &=& R + 2d D_{a}\partial^{a}\beta +
d(d+1)\partial_{a}\beta\partial^{a}\beta
            - d(d -1)k_{2}e^{-2\beta} \:, \\
\hat{G}_{ab} &=& G_{ab} + t_{ab}   \label{gab},  \\
\hat{G}_{mn} &=& \frac{1}{2}\bar{g}_{mn} [ \: R +
2(d -1)D_{a}\partial^{a}\beta
            + d(d -1)\partial_{a}\beta\partial^{a}\beta   \nonumber \\
               &   & - (d -1)(d -2)k_{2}e^{-2\beta} \: ] \: .
\label{gmn}
\end{eqnarray}
Here the generalized energy momentum tensor $t_{ab}$ is given by
\begin{eqnarray}
t_{ab} & \equiv & \frac{1}{2}g_{ab} [ \: 2d D_{c}\partial^{c}\beta
                  + d (d +1)\partial_{c}\beta\partial^{c}\beta
                  - d (d -1)k_{2}e^{-2\beta} \: ]               \nonumber
\\        &        & - d( \: D_{a}\partial_{b}\beta
                  + \partial_{a}\beta\partial_{b}\beta \: ) \: .
\end{eqnarray}
Note that $\hat{g}^{MN}\hat{G}_{MN}= (\frac{D}{2} -1)\hat{R}$. Hence
one can obtain the following equation of $\varphi$
\begin{equation}
\partial_a \varphi \partial^a \varphi + D_a \partial^a \varphi + d
\partial_a \beta \partial^a \varphi
= \kappa_3 \,\,
 \frac{e^{- \varphi}}\epsilon [ (D -2) \partial_\varphi U
- D U]    \label{phi3}
\end{equation}
from eliminating the $\hat{R}$ term in the trace of $\hat{G}_{MN}$ equation
(\ref{gmnd}) and the $\varphi$ equation (\ref{phi2}).
Here one has set $\kappa_3= 4 \epsilon /[4(D-1) \epsilon +D-2]$.
Finally, one can show that the trace of the $\hat{G}_{mn}$
equation (\ref{gmn2}),
$\bar{g}^{mn}\hat{G}_{mn}$, and the trace of the $\hat{G}_{mn}$
equation (\ref{gmn}) gives two constraint-equations
related to $R$. One can eliminate these $R$ terms
and obtain the following equation
for $b(t)$:
\begin{eqnarray}
d \partial_a \beta   \partial^a \beta   + D_a \partial^a \beta - (d -1)
k_2 e^{- 2 \beta}
&=& (1+ {1 \over 4 \epsilon}) [
\partial_a \varphi \partial^a \varphi + D_a \partial^a \varphi + d
\partial_a \beta \partial^a \varphi ]  \nonumber \\ & &
- \partial_a \beta \partial^a \varphi
+ \frac{e^{- \varphi}}\epsilon [ U- \partial_\varphi U]
 \label{beta3}
\end{eqnarray}
Therefore we will take (\ref{beta3}) as the independent $\beta$-equation.
This will soon be shown to be helpful in our analysis below. 
Finally, one can show that $G_{tt}$ component of the equation (\ref{gab2}), the
$\varphi$ equation (\ref{phi3}) and the $\beta$ equation (\ref{beta3})
becomes
\begin{eqnarray}
{\alpha}'^2 +\frac{k_1}{a^2} +d {\alpha}' {\beta}'
+ \frac{d (d -1)}{6} ( {\beta}'^2 + k_2 e^{-2 \beta})
&+& {\alpha}'{\varphi}' +\frac{d}{3}{\beta}' {\varphi}'
= {1 \over 24 \epsilon}{\varphi}'^2 +
\frac{U}{3\epsilon}  e^{-\varphi}  \label{alphaf} \\
\varphi'' +3 \alpha' \varphi' + d \beta' \varphi' + \varphi'^2
&=&
- \frac{\kappa_3}{\epsilon} e^{- \varphi} [ (D -2) \partial_\varphi U - D
U] , \label{phif} \\
\beta'' + 3 \alpha' \beta' +d \beta'^2 + (d -1) k_2 e ^{-2\beta}
+\beta' \varphi' &=&
\frac{\kappa_3}{\epsilon} e^{-\varphi} [ \partial_\varphi U + (1+ {1 \over 2 \epsilon} )U]  .
\label{betaf}
\end{eqnarray}
Note also that the $G_{ij}$ component of the equation (\ref{gab2}) can be
deduced from the $4$-dimensional Bianchi Identity $D_aG^{ab}=0$ associated with
the $4$-dimensional FRW metric. Hence it is in fact redundant. 
Therefore, equations (\ref{alphaf}-\ref{betaf}) are in fact a complete
set of equations of motion one needs for solving $\alpha$, $\beta$ and $\varphi$.

\section{Inflationary Universe}

If one assumes the slow-rollover approximation, namely,
$\frac{a'}{a} \gg |\varphi'|$,
one can show that
\begin{eqnarray}
{\alpha}'^2  +d {\alpha}' {\beta}'
+ \frac{d (d -1)}{6} {\beta}'^2
&=&
\frac{U}{3\epsilon}  e^{-\varphi} \label{alpha} ,\\
 3 \alpha' \beta' + d  \beta'^2
&=& \frac{\kappa_3}{\epsilon} e^{-\varphi} [
\partial_\varphi U +(1+{1 \over 2 \epsilon}) U] , \label{beta} \\
(3 \alpha' +d \beta') \varphi' &=&
 \frac{\kappa_3}{\epsilon} e^{- \varphi} [(d+4)U -(d +2) \partial_\varphi
U] . \label{varphi}
\end{eqnarray}
Here we have set $k_1=k_2=0$ for simplicity.
Note that the issue of the non-compact internal space has recently been subject of renewed 
interest \cite{visser}. 
An exotic class of Kaluza-Klein models in which the internal space is neither
compact nor even of finite volume was considered and  
Gravity is used to trap particles near a four-dimensional subminifold
of the higher dimensional spacetime. 

Moreover, we have also assumed that $|\varphi ''| \ll \alpha' |\varphi '|$
and $|\beta ''| \ll \beta'^2$. We will show shortly that these assumptions
can be met rather easily.

We will assume for the moment during the slow-rollover period that
$\alpha = \alpha_0 t$ and $\beta = -k \alpha_0 t$ for some positive
real number $k$ and $\alpha_0$. This kind of solution represents a 
brief moment of inflating scale factor $a$ accompanied by a contracting internal 
scale factor $b$. This will be helpful in finding possible constraint on the 
form of symmetry breaking potential one would require for a more realistic model.
One can also assume that $U \sim U_0 \equiv V(\phi=\phi_0) $ while $\phi \sim \phi_0$ during the
inflationary phase. Therefore, one can show that equations (\ref{alpha}-\ref{varphi})
 can be brought to the following form:
\begin{eqnarray}
d(d-1)k^2 -6 dk +6 
&=& \tilde{k}
 \label{talpha} ,\\
 k(dk-3)
&=& {\tilde{k} \kappa_3 \over 4} (s -s_- ) \label{tbeta} , \\
dk-3 &=& {(d+2) \tilde{k} \kappa_3 \alpha_0 \over 4 \varphi_0'} (s -s_+ ).
 \label{tvarphi}
\end{eqnarray}
Here $\tilde{k} \equiv 2 U_0 /\epsilon \alpha_0^2 \phi_0^2$, 
$\varphi_0' \equiv \varphi'(t_0)$,
$s_- \equiv -2 - 1/ \epsilon $ and $s_+ \equiv 2 (d+4)/(d+2)$.
In addition, we have also defined $s \equiv \phi_0 (\partial_\phi U)_0
/U_0$ as the scaling factor of $U$ evaluated at $\phi=\phi_0$.

We will first study the case where $d > 1$.
Note that equation (\ref{talpha}) indicates that $d(d-1)k^2 -6 dk +6 
\equiv d(d-1)(k-k_+)(k-k_-)>0$ under the assumption 
that $U$ is positive everywhere. Here we have defined $k_{\pm} \equiv
3/(d-1) \pm \sqrt{3d(d+2)} /d(d-1)$ as the roots of the $k$-equation.
Therefore one has either $k>k_+$ or $k< k_-$ from the $k$-inequality.
In addition, equation (\ref{tbeta}) and equation (\ref{tvarphi}) gives 
\begin{equation}
\alpha_0 = {s -s_- \over k(d+2) (s-s_+)} \varphi_0' \label{alphaphi}.
\end{equation} 
This shows that 
$\varphi_0' (s-s_+)(s-s_-) >0$ since $k$ is assumed to be positive. 

Moreover, one also assumes that $dU(\phi)/dt <0$ such that the scalar field is rolling down
from some initial value $\phi_0$ to the minimum potential energy state $\phi_m$.
This means that $s\varphi_0' <0$. Therefore, one finds that there are only
two kinds of combination capable of supporting this process.
The first one is (1) $\varphi_0' <0$, $0< s <s_+$ and the second one is
(2) $\varphi_0' >0$, $s >s_-$. One can further rule out case (2) from the assumption
that $\alpha_0 \gg |\varphi_0'|$. Indeed, equation (\ref{alphaphi}) indicates
that the slow-rollover assumption is equivalent to
$|s -s_- | \gg  k(d+2) |s-s_+|$. Hence case (1) can be shown to give a constraint
\begin{equation}
s \gg {k(d+2) s_+ +s_- \over k(d+2) +1}. \label{sconst}
\end{equation}
This can easily be achieved provided that $s_+ \gg s_-$. 
Note that this is true if $\epsilon \ll 1$ \cite{acc}.
Similarly, case (2) will give a contradictory result $s_- \gg s_+$.
Therefore, case (2) is ruled out.

In addition, case (1) and equation (\ref{tvarphi}) shows that $dk-3>0$. Hence 
the constraint on $k$ obtained earlier is further restricted
to the case where  $k>k_+$. This is because
$3/d<k<k_-$ leads to a contradiction $3 > d (d+2)$.

In short, the induced Kaluza-Klein compactification
admits chaotic inflation only if the symmetry breaking potential obeys
a number of constraints listed earlier.
They are :
\begin{eqnarray}
 &(a)& \: s_+ \gg s_- , \label{conda} \\ 
&(b)& \: s_+ > s \gg [k(d+2) s_+ +s_-] / [ k(d+2) +1],  \label{condb}  \\
&(c)& \: \varphi'_0 <0, \label{condc}  \\
&(d)& \: k > k_+ .  \label{condd} 
\end{eqnarray}
For example, one would have
(a) $8/5 \gg s_-$, (b) $8/5 > s \gg (5k +s_-)/ (8k+1)$ and
(c) $k>1$ as the constraint on $k$ and $s$ for the case where $d=6$ 
or equivalently $D=10$.

One can easily construct an effective symmetry breaking potential by
expanding the potential around the initial point $\phi_0$.
Explicitly, it will take the following form $U=U_0 + sU_0 (\phi-\phi_0) + \cdots$
around the initial point. For example, one can show that the conventional $\phi^4$ model with
$U=(\lambda/8) (\phi^2 -v^2)^2$ does not satisfy the constraint obtained earlier.
Indeed, one can show that equation (\ref{tvarphi}) gives 
\begin{equation}
\Lambda > {\lambda \over 8(d+4)} (\phi_0^2 -v^2) [d \phi_0^2 +(d+4)v^2]
\label{Lambda} 
\end{equation}
for the conventional $\phi^4$ model with an additional positive definite cosmological constant term 
$\Lambda$.
This clearly shows that the chaotic inflation condition 
$\phi_0^2 > v^2$ is inconsistent with the case where 
$\Lambda = 0$. Note that the no-hair conjecture states that
cosmologies with a positive cosmological constant would approach the de Sitter solution asymptotically \cite{ghm}.
Even some counter examples are found, it was shown to hold for a very general conditions \cite{JSS}.
Our result appears to favor above conjecture with the inclusion of the higher dimensional space.
Therefore, the conventional $\phi^4$ model with 
vanishing cosmological constant can not support an 
inflationary solution with expanding external-space 
and contracting internal-space. We will solve the conventional $\phi^4$ model later in section IV.

For the case where $d=1$, the situation is rather different.
Equation (\ref{talpha}) implies that $k<1$ for positive $U_0$.
In addition, equation (\ref{tbeta}) gives $s< s_- (<0)$ while
equation (\ref{tvarphi}) implies $\varphi'_0 >0$ (new inflationary
solution). In addition, the slow-rollover assumption indicates that
$s_- -s \gg k (10-3s)$. Therefore one obtains $(3k-1)s \gg 10k-s_-(>0)$. 
This implies that $k < 1/3$. In summary, one has
(a) $s< s_- (<0)$, (b) $k < 1/3$, and (c) $\varphi'_0 >0$.
Therefore the five dimensional Kaluza-Klein new inflationary solution
with expanding external-space and contracting internal-space
can also be arranged if the field parameters are chosen appropriately.

One can also study the case where the internal-space scale factor remains constant, $i.e.$
$b=b_0$ \cite{kal} or equivalently $k=0$ in the early universe.
In this case, the equations will become
\begin{eqnarray}
{\alpha}'^2 +\frac{k_1}{a^2}
+ \frac{d (d -1)}{6}  {k_2 \over b_0^2}
&+& {\alpha}'{\varphi}'
= {1 \over 24 \epsilon} {\varphi}'^2 +
\frac{U}{3\epsilon}  e^{-\varphi}  \label{alphafb} , \\
\varphi'' +3 \alpha' \varphi'  + \varphi'^2
&=&
- \frac{\kappa_3}{\epsilon} e^{- \varphi} [ (D -2) \partial_\varphi U - D
U] , \label{phifb} \\
 (d -1) {k_2 \over b_0^2}
 &=&
\frac{\kappa_3}{\epsilon} e^{-\varphi} [ \partial_\varphi U + ({1 \over 2 \epsilon} +1)U]
\label{betafb}   .
\end{eqnarray}
Therefore one finds that there is a strong constraint (\ref{betafb}) left over
for the $b$ equation. This equation says that
$\partial_\varphi U + (1+{1 \over 2 \epsilon} )U=0$ for a flat 
internal-space (i.e. $k_2=0$). One can then show that either (i) the potential $U$
has to be a special
fractional polynomial functional of $\phi$, namely, $U=k_0 \phi^{-(2+1/\epsilon)}$ with
a proportional constant $k_0$, or (ii) the dynamics
of the scalar field has to be frozen, namely, the scalar field
becomes a constant $\phi=\phi_0$.
One can show that the first case would imply that $\alpha' \sim \varphi'
/2$ under the constraint $\epsilon \ll 1$.
This contradicts
the slow rollover approximation. On the other hand, the case (ii)
implies that $U(\varphi_0) = \partial_\varphi U(\varphi_0)=0$. Hence one has
$a'^2=-k_1$ due to equation (\ref{alphafb}). Therefore, one needs
$k_1=-1$ in order to admit a power law
inflation. One can hence tune the field parameters to induce enough
inflation with expanding external-space and constant internal-space. 
But this model can not tell us when the inflationary
phase should come to an end. One would have to expect that this induced gravity
model remains valid only during the inflationary period  and leave the 
problem to other resolutions.

On the other hand, one can show that the constraint (\ref{betafb})
\begin{equation}
 (D -5) {k_2 \epsilon \over b_0^2 \kappa_3}  \phi^2
=
 \partial_\varphi U + (1+ {1 \over 2 \epsilon} )U      \label{k2u}
\end{equation}
implies
$\phi=\phi_0$ for the case $k_2 \ne 0$ unless
\begin{equation}
U=k_0 \phi^{-2-{1 \over 4 \epsilon}} + {2(D -5) k_2 \epsilon^2 \over 
(1+ 4 \epsilon) b_0^2 \kappa_3} \phi^2 . \label{nobu}
\end{equation}
If $\phi=\phi_0$, equation (\ref{phifb}) implies that
\begin{equation}
\left[ (D-2) \partial_\varphi U -DU \right]_{\phi_0}=0 . \label{nobb}
\end{equation}
Equations (\ref{k2u}) and (\ref{nobb}) mean that all field parameters and initial
conditions are constrained by these equations.
In addition, equation
(\ref{alphafb}) tells us that
\begin{equation}
\alpha'^2 \sim   (D -5) {k_2 \over 3 b_0^2 }
\end{equation}
independent of the form of potential $U$. Of course, the initial value of the
scalar field $\phi_0$ is determined by the form of potential and the two
constraints just derived.
This solution is an inflationary solution with expanding external-space and constant 
internal-space as long as $k_2 (D-5) >0$ and $b_0
\ll 1$. One can certainly tune $b_0$ to induce enough inflation 
with expanding external-space and constant internal-space.
But this solution can not tell us how to exit the inflationary phase at
this
point either. One would then have to expect again 
that this kind of induced gravity model
would not remain effective as soon as the inflationary process is completed.

On the other hand equations (\ref{alphafb}) and (\ref{phifb}) imply that
\begin{eqnarray}
\alpha'^2 &=&  (1 + 6 \epsilon) A +B , \\
\alpha'\varphi' &=& 4 \epsilon A + 2 B ,
\end{eqnarray}
under the slow-rollover approximation if $U$ is given by equation (\ref{nobu}).
Here
\begin{eqnarray}
A&=& {(D-5) k_2 \over 3 b_0^2 (1 + 4 \epsilon)}, \\
B&=& { k_0 \over 3 \epsilon \phi_0^{4+ {1 \over 4 \epsilon}}}
\end{eqnarray}
for $\phi \sim \phi_0$ in this inflationary phase. Therefore
one can choose $\epsilon \ll 1$ and $B \ll A$ in order to be consistent with
the assumption that $\alpha' \gg |\varphi'|$. In addition, one can choose
$A>0$ since $\alpha'^2 >0$. This implies that $k_2=1$ for $D \ge 5$. In
addition, $A \gg B$ implies that
$k_0 b_0^2 \ll (D-5) \epsilon \phi_0^{4+ {1 \over 4 \epsilon}}$
which can be achieved by tuning the field parameters appropriately.
Moreover, one still needs to make sure that the potential $U$ given by equation
(\ref{nobu}) has at least a local minimum $\phi_m$ far away from the initial
data $\phi_0$ such that inflation can exit in due time.

Fortunately, a local minimum always exists for a large class of parameters.
Indeed, one can show that
\begin{equation}
k_0b_0^2 \sim (D-5)(D-2) \epsilon^2
\phi_m^{4+ {1 \over 4 \epsilon}} 
\end{equation}
from $\partial_\phi U |_{\phi=\phi_m}=0$. Hence one only needs
\begin{equation}
\epsilon \phi_m^{4+ {1 \over 4 \epsilon}}  \ll
{2 \over D-2} \phi_0^{4+ {1 \over 4 \epsilon}}. 
\end{equation}
In addition, the requirement $U''|_{\phi_m} >0$ can be made valid very easily.

Therefore, the inflationary process can properly work with the assumption
$b=b_0$ for the case where $F^{d} =S^{d+1}$. And this has to come along with the
potential of the form given by equation (\ref{nobu}). 

\section{Conventional $\phi^4$ Model}

One can also works on the model with a spontaneously symmetry breaking (SSB) $\phi^4$ potential $U=\frac{\lambda}8 
(e^\varphi -v^2)^2$. This sort of potential will be referred to as the conventional $\phi^4$ model
in this paper. 
It is straightforward to show that
$\partial_\varphi U= \frac{\lambda}4 (e^\varphi -v^2)e^\varphi$.
Hence equation (\ref{varphi}) becomes
\begin{equation}
(3 \alpha' +d \beta') \varphi' =
 -\frac{\kappa_3\lambda}{8\epsilon} e^{- \varphi}
 (e^\varphi -v^2)[de^\varphi +(d+4)v^2] \label{tovac} .
\end{equation}
This indicates that  $3 \alpha +d \beta$ is always an increasing function
as long as the $\varphi$ field is rolling down to its true vacuum
$e^\varphi = v^2$.
It also indicates that $\phi$ cannot go far away from its local minimum,
hence it should oscillate around $\phi=v$ after the inflation is over.
We will come back to this point shortly.
Moreover, the equation (\ref{beta}) becomes
\begin{equation}
 3 \alpha' \beta' + d  \beta'^2
= \frac{2U}{(d+2)\epsilon} e^{-\varphi}        \label{betap}
\end{equation}
if $\epsilon \ll 1$ and $|\phi^2-v^2|/\phi^2 \gg 4 \epsilon$. These assumptions
can be adjusted rather easily.
Together with equation (\ref{alpha}), one finds that
\begin{equation}
2\alpha'^2 +(d - 2) \alpha' \beta' -d \beta'^2
= (\alpha' -\beta')(2 \alpha' +d \beta') \sim 0. \label{higgsl}
\end{equation}
This means that $\alpha'= \beta'$ because the equation $\alpha'=- {d \over 2} \beta'$
contradicts the equation (\ref{betap}). Hence $b(t)$ increases along with the
expanding $a(t)$ in the inflationary era under the slow-rollover
approximation. Therefore, one has shown that the conventional $\phi^4$ model supports DFRW space
instead of the 4dFRW space. Hence the solution with expanding external-space and
contracting internal-space can not be found under the slow-rollover approximation.
We will still, however, study the DFRW solution in details in this section for
completeness.
Note that the presence of a nonvanishing cosmological constant in the conventional  $\phi^4$ model
will not affect the equation (\ref{higgsl}) under the same slow rollover approximation.

Note that equation (\ref{betap}) gives us
\begin{eqnarray}
\alpha' & \sim & \sqrt{ { \lambda v^4 \over 8 (d +2)(d +3) } } \,\,\,
{ v \over \phi_0 } , \label{alphas}  \\
a & \sim & a_0 e^{ \sqrt{ { \lambda v^4 \over 8 (d +2)(d +3) } }
       \,\,\,\,   { v \over \phi_0 } t} \label{as}       .
\end{eqnarray}
Here one has set ${1 \over 2} \epsilon v^2=1$ such that the gravitational
constant measured today is set as $1$ in Planck unit. Moreover,
$\phi_0$, set to be positive, denotes the initial value of $\phi$ field.
Moreover, equation (\ref{tovac}) gives
\begin{eqnarray}
\phi' & \sim & \sqrt{ { \lambda (d +4)^2 \over 2 (d +2)(d +3) } } \,\,\,
v  , \label{phis}  \\
\phi & \sim & \phi_0 + \sqrt{ { \lambda (d +4)^2 \over 2 (d +2)(d +3) } }
\,\,\, v t    \label{phit} .
\end{eqnarray}
Here we can see that the assumptions
$|\varphi ''| \ll \alpha '|\varphi '|$
and $|\beta ''| \ll \beta'^2$ are both satisfied without imposing 
any further constraints.

One can further derive a few inequalities from the slow-rollover
assumption $\frac{a'}{a} \gg |\varphi'|$.
First of all, they give
\begin{equation}
v^2 \gg 4 (d +4) \label{ieone} .
\end{equation}
Note that the cosmological constant term $-{1 \over 8} \lambda v^4$ at initial
time
should be less than $1$, in Planck unit, in order that quantum effect
can be neglected.
We will be using Planck unit from now on.
In addition, if the scale factor $a(t)$ is capable of expanding some $60$ e-fold 
in a time interval of roughly $\Delta T \sim 10^8$ Planck unit, one should
have the following inequality:
\begin{equation}
 {\lambda v^4 \over 8} \ge
 (d +2)(d +3) {\phi_0^2 \over v^2}   \times 3.6 \times 10^{-13} .
\label{ietwo}
\end{equation}
Inequality (\ref{ietwo}) can be made valid rather easily.
Indeed, these inequalities can be easily satisfied by choosing
large $v^2$ (hence small $\epsilon$) and a $\lambda$ around the order of
$10^{-17}$ as in Ref. \cite{acc}.
Hence one shows that the slow-rollover approximation is indeed a good
approach to this expanding solution.

Note that we can also extract information about $\alpha$, $\beta$
and $\varphi$ when $\varphi \to v$ near the end of the expansion. 
This can be done by analyzing
equations (\ref{alphaf}-\ref{betaf}) by assuming $e^\varphi = \xi +v^2$
with $\xi \ll v^2$.
Moreover, one can show that equations
(\ref{phif}-\ref{betaf}) become,
\begin{eqnarray}
\partial_t (e ^{3 \alpha+d \beta + \varphi} \varphi') &=& - {\kappa_3 \lambda
\over 8 \epsilon} e^{3 \alpha + d \beta} ( e^{\varphi} -v^2 )
[d e^\varphi + (d +4)v^2 ], \label{phiad} \\
\partial_t (e ^{3 \alpha+d \beta + \varphi} \beta') &=& {\kappa_3 \lambda
\over 8 \epsilon} e^{3 \alpha + d \beta} ( e^{\varphi} -v^2 )
[(3 +{1 \over 2 \epsilon} ) e^\varphi - (1 +{1 \over 2 \epsilon} )v^2 ]. \label{betaad}
\end{eqnarray}
Hence equations (\ref{phiad}-\ref{betaad}) become
\begin{eqnarray}
\partial_t (e ^{3 \alpha+d \beta } \xi') &=& -
m_v (d +2)  e^{3 \alpha + d \beta} \xi       ,
\label{phiv} \\
\partial_t [e ^{3 \alpha+d \beta } \beta'(\xi+v^2) ] &=&
m_v  e^{3 \alpha + d \beta} \xi  \label{betav}  ,
\end{eqnarray}
as one takes the limit $e^\varphi = \xi +v^2$.
Here $m_v \equiv \kappa_3 {\lambda v^4 \over 8}$.

Moreover, equations (\ref{alphaf}-\ref{betaf}) can be interpreted as a
set of equations that allow one to express $\alpha(t)$ and $\beta(t)$
as functions of $\xi(t)$. Therefore one can expand
$\alpha$ and $\beta$ as polynomials of $\xi$, i.e.
one can write
\begin{eqnarray}
\alpha (t) &=& \alpha_0(t) + \alpha_1(t) \xi(t) + \alpha_2(t) \xi^2 (t) +
\cdots \label{alphaexp}  \\
\beta (t) &=& \beta_0(t) + \beta_1(t) \xi(t) + \beta_2(t) \xi^2 (t) +
\cdots . \label{betaexp}
\end{eqnarray}
Therefore, the lowest (first) order in $\xi$ of equation (\ref{phiv})
is
\begin{equation}
\partial_t (e ^{3 \alpha_0 +d \beta_0 } \xi') = -
m_v (d +2)  e^{3 \alpha_0 + d \beta_0} \xi .
\label{phiv1}
\end{equation}
Moreover, the zeroth order in $\xi$ of equation (\ref{betav}) can be shown to
be
\begin{equation}
 \partial_t [e ^{3 \alpha_0+d \beta_0 } \beta_0'] =0.
\label{betav0}
\end{equation}
This means that $\partial_t [e ^{d \beta_0 }] ={\rm const.}\times e ^{-3
\alpha_0}$. Therefore, one has $\partial_t [e ^{d \beta_0 }] \sim 0$ since $e
^{-3 \alpha_0}$ is very closed to $0$ in the post-expansion era.
Therefore, one can assume that $e ^{d \beta_0 }$ is changing very slowly
as $\phi^2 \to v^2$. In addition, the zeroth order in $\xi$ of equation
(\ref{alphaf}) is
\begin{equation}
\alpha_0'^2 + d \alpha_0' \beta_0 ' + {d^2 - d \over 6} \beta_0'^2
\sim 0   . \label{alphav0}
\end{equation}
This gives
\begin{equation}
 \alpha_0' = - { \sqrt{3} d \pm \sqrt{d^2 +2 d}
\over 2 \sqrt{3} } \beta_0' \sim 0  \label{alphavv}
\end{equation}
to this order of limit. Hence equation
(\ref{phiv1}) becomes an equation for a simple harmonic oscillator
\begin{equation}
\xi '' = - m_v (d +2) \xi . \label{sho}
\end{equation}
Note that the left hand side of equation (\ref{sho}) approaches
$- \lambda v^2 \xi$ in the limit $\epsilon \ll 1$.
In short, one finds that $\phi$ field indeed oscillates about the local minimum
of the symmetry breaking potential $U$.
Furthermore, equations (\ref{alphavv}) indicates that $\alpha_0' \beta_0' < 0$
as $\phi$ approaches the local minimum of $U$.
Therefore, $b(t)$ in fact starts decreasing if $a(t)$ remains increasing at later
time. Note that above analysis is only a rough estimate, but it gives us a
rough picture of what is going on when $\xi$ field approaches zero.

\section{Conclusions}

In summary,
a D-dimensional induced gravity model in 4dFRW space is studied carefully.
We present a careful and detailed analysis for the compactification
process. This model is then solved for the inflationary solution
in the slow-rollover approach. 
A number of constraints on the symmetry breaking
potential are found. These constraints are derived from the search for a inflationary
solution with expanding external-space and contracting, compactified 
internal-space.
 
The result indicates that the possible form of symmetry breaking
potential, prescribed by $s$, 
is constrained by equations (\ref{talpha}-\ref{tvarphi}) due to the field equations.
Here, $s \equiv \phi_0 (\partial_\phi U)_0
/U_0$ signifies the scaling factor of $U$ evaluated at $\phi=\phi_0$.
The cases where $d>1$ and $d=1$ are analyzed separately. Explicitly,
constraints to the coupled potential
are listed in equations (\ref{conda}-\ref{condd}) for the case
where $d>1$. In particular,  one shows
that these constraints read 
(a) $8/5 \gg s_-$, (b) $8/5 > s \gg (5k +s_-)/ (8k+1)$ and
(c) $k>1$ in the limit where $d=6$.
It was then shown that the conventional $\phi^4$ model with an additional cosmological constant term fails to
satisfy the above constraints. On the other hand, one shows that
(a) $s< s_- (<0)$, (b) $k < 1/3$, and (c) $\varphi'_0 >0$ for the case where $d=1$.
In addition, we
also solve the case where the internal scale factor $b$ remains constant during
the inflationary phase.

An expanding solution is also found and analyzed for the conventional $\phi^4$ model.
In order to generate a solution with expanding external-space inflation in the very 
early universe, one finds that the internal space is expanding too under the 
slow rollover approximation. Therefore, this indicates 
that dimensional reduction has to be completed before expanding external-space starts to expand. 
With properly chosen free parameters and boundary conditions of the scalar
field, one shows that enough expansion can be easily achieved regardless of the
negative impact of the expanding internal-space in the conventional $\phi^4$ model.

{\bf Acknowledgments :}
This work is supported in part by the National Science Council under
the contract number NSC88-2112-M009-001.

\end{document}